# Colossal Effect of Nanopore Surface Ionic Charge on the Dynamics of Confined Water

Armin Mozhdehei[1], Philip Lenz[2,3], Stella Gries[4,8], Sophia-Marie Meinert[2], Ronan Lefort[1], Jean-Marc Zanotti[5], Quentin Berrod[6], Markus Appel[7], Mark Busch[4,8], Patrick Huber[4,8,*], Michael Fröba[2,3,*], Denis Morineau[1,*]

1.  Institute of Physics of Rennes, CNRS-University of Rennes, UMR 6251, F-35042 Rennes, France.
2.  Institute of Inorganic and Applied Chemistry, University of Hamburg, 20146 Hamburg, Germany.
3.  The Hamburg Centre for Ultrafast Imaging, Luruper Chaussee 149, 22761 Hamburg, Germany
4.  Institute for Materials and X-ray Physics, Hamburg University of Technology, Hamburg 21073, Germany
5.  Université Paris-Saclay, Laboratoire Léon Brillouin, CEA, CNRS, F-91191 Gif-sur-Yvette, France
6.  Univ. Grenoble Alpes, CNRS, CEA, Grenoble INP, IRIG, SyMMES, F-38000 Grenoble, France
7.  Institut Laue-Langevin, F-38042 Grenoble, France
8.  Centre for X-ray and Nano Science CXNS, Deutsches Elektronen-Synchrotron DESY, 22603 Hamburg, Germany.

Email addresses : *patrick.huber@tuhh.de, *michael.froeba@uni-hamburg.de, *denis.morineau@univ-rennes.fr

**TOC Graphic**

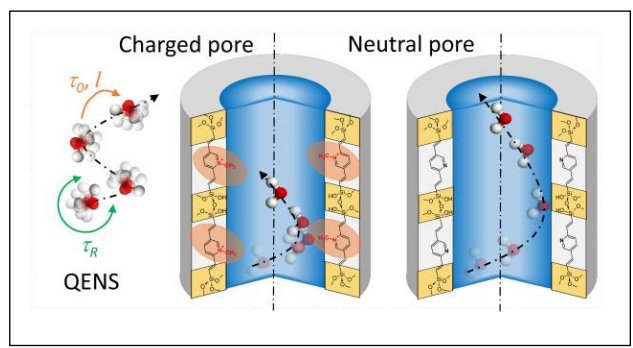

Sketch of rotational and translational motions of water, as well as the molecular structure of the pores.

# Abstract

Interfacial interactions significantly alter the fundamental properties of water confined in mesoporous structures, with crucial implications for geological, physicochemical, and biological processes. Herein, we focused on the effect of changing the surface ionic charge of nanopores with comparable pore size (3.5-3.8 nm) on the dynamics of confined liquid water. The control of the pore surface ionicity was achieved by using two periodic mesoporous organosilicas (PMOs) containing either neutral or charged forms of a chemically similar bridging unit. The effect on the dynamics of water at the nanoscale was investigated in the temperature range of 245 – 300 K, encompassing the glass transition by incoherent quasielastic neutron scattering (QENS), For both types of PMOs, the water dynamics revealed two distinct types of molecular motions: rapid local movements and translational jump diffusion. While the neutral PMO induces a moderate confinement effect, we show that the charged PMO drastically slows down water dynamics, reducing translational diffusion by a factor of four and increasing residence time by an order of magnitude. Notably, by changing the pore filling values, we demonstrate that for charged PMO this effect extends beyond the interfacial layer of surface-bound water molecules to encompass the entire pore volume. Thus, our observation indicates a dramatic change in the long-range character of the interaction of water confined in nanopores with surface ionic charge compared to a simple change in hydrophilicity. This is relevant for the understanding of a broad variety of applications in (nano)technological phenomena and processes, such as nanofiltration and membrane design.

# 1. Introduction

Water is fundamental to life, and albeit a simple molecule, it exhibits unique and complex behavior that remains a subject of intense scientific investigations. Water's remarkable properties, such as its high heat capacity, anomalous density behavior, and complex hydrogen-bonding network, become even more pronounced when it is confined within small pores.[1-3] In this regard, the behavior of water confined in mesoporous materials has been extensively studied due to its implications in numerous scientific and industrial fields, such as capillarity-driven water transport,[4] catalysis,[5-7] energy storage,[8-11] nanofiltration,[12] and environmental remediation.[13] Unlike bulk water, water confined in nanoscale environments exhibits unique structural and dynamical properties due to spatial constraints, surface interactions, and changes in hydrogen-bond networks.[14-17] Studies by Gallo et al. and Debenedetti and Stillinger have shown that these properties are significantly altered under confinement, where the interactions between water molecules and the surrounding surfaces lead to deviations from bulk behavior.[18, 19] In confined systems, water can form distinct layers with different structural and dynamical characteristics depending on its proximity to the pore walls. The molecules closest to the surface often exhibit reduced mobility due to stronger hydrogen-bonding interactions with surface functional groups, while molecules located toward the center of the pore may retain more bulk-like behavior.[20, 21] As a whole, these studies indicate that the influence of nanoporous confinement on the water properties arise from the complex interplay of different effects, such as low-dimensionality, spatial restriction, and surface interaction.

Mesoporous silica materials, such as MCM-41 and SBA-15, and periodic mesoporous organosilicas (PMOs)[22] provide ideal model systems for exploring how confinement and surface chemistry (i.e., surface charge, surface polarity, hydrophilicity/hydrophobicity, and so forth) influence water dynamics. These materials are characterized by their tunable pore sizes, well-defined geometries, and versatile surface chemistries, allowing researchers to systematically study the impact of surface interactions on confined water.[23-28] Jani et al., by doing quasi-elastic neutron scattering (QENS) measurements, demonstrated that modifying the hydrophilicity of the pore walls can significantly influence water mobility. In hydrophilic MCM-41, water molecules form strong hydrogen bonds with silanol groups on the surface, resulting in slower diffusion and rotational dynamics. Conversely, hydrophobic PMOs, which contain organic bridging units like benzene, reduce surface-water interactions, allowing water to behave more similarly to its bulk counterpart.[25] Moreover, Malfait et al. further expanded on these findings by investigating the role of surface chemistry in shaping the rotational dynamics of water confined in PMOs. Using techniques such as Raman spectroscopy and dielectric relaxation spectroscopy, they revealed that water near hydrophilic surfaces tends to form a more structured hydrogen-bond network, whereas water near hydrophobic surfaces is less constrained and thus more dynamic.[29, 30]

The complexity of water behavior in bulk and confined spaces increases when ions are present.[23, 26, 31, 32] Previous studies have explored this phenomenon in aqueous electrolytic solutions confined within porous silica. The researchers demonstrated how cation-silica surface interactions affect water dynamics, finding that the type of ion (e.g., kosmotropic vs. chaotropic) plays a significant role in determining water mobility within nanoconfined systems. Kosmotropic ions, such as $Li^+$, enhance water structuring near surfaces, thereby reducing diffusion rates, whereas chaotropic ions like $Cs^+$ have a lesser impact on water dynamics, allowing for faster molecular motion. These findings have significant implications for optimizing technologies like energy storage in batteries, fluid transport in nanofiltration or electrosorption-induced actuation,[33, 34] where electrolyte behavior under confinement is critical for performance.

Building on these foundational studies, the present work addresses a focused version of a broader question: how does confinement within ionically charged porous media influence water dynamics? To explore this in a controlled way, we use a model system of periodic mesoporous organosilicas (PMOs) with identical pore structures but with either neutral or ionically charged inner surfaces. Using quasielastic neutron scattering (QENS), we systematically probe the translational dynamics of water confined in these two materials. Our findings reveal a pronounced effect of surface charge on water mobility, even beyond the 2-3 interfacial layers. Although limited to a single specific ionically charged PMO system, these results significantly expand the current understanding of confined water, which has so far been mainly based on neutral hydrophilic or hydrophobic surfaces.

## 2. Materials and methods

### 2.1. Samples

Two different PMO powders with comparable porous geometry but different surface chemistry were prepared. The first one, referred as DVP-PMO, is a divinyl-pyridine bridged PMO, which presents an essentially neutral pore surface. The second one, referred as DVMeP-PMO, is a divinyl-methyl pyridinium bridged PMO, which presents a positively charged pore surface induced by the presence of pyridinium cations. This charged porous material was derived from the post-synthetic modification of DVP-PMO using methyl iodide (MeI).

The structure of the two PMOs can be described as a regular hexagonal lattice of parallel and cylindrical channels, conforming a honeycomb structure like MCM-41 porous silicas, as schematized in Fig. 1. In contrast, the pore walls of the PMO's feature organic units bridging inorganic silica units, which alternate periodically along the main pore axis. The molecular-scale periodicity of the arrangement of organic/silica units along the pore axis is demonstrated by (00l) Bragg reflections, while, like for MCM-41, the hexagonal in-plan arrangement of porous channels is demonstrated by (hk0) Bragg reflections.

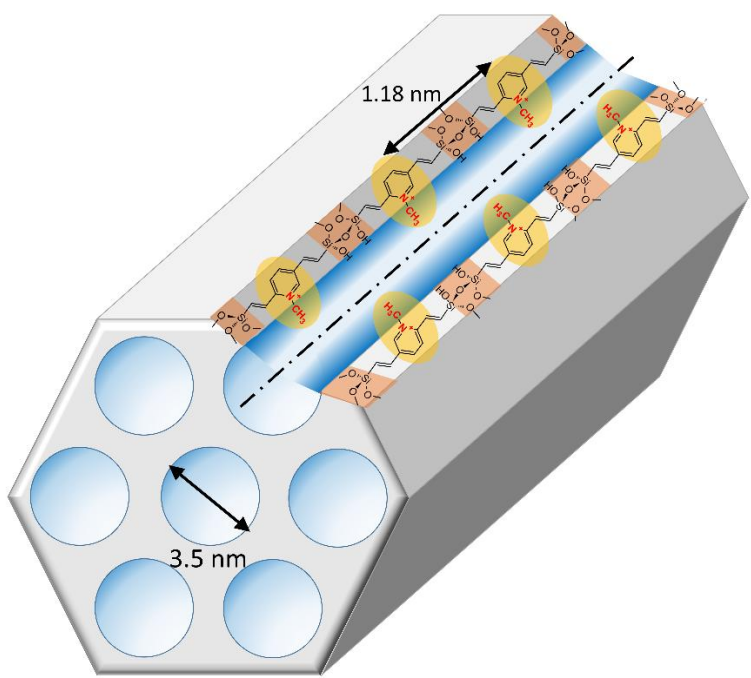

*Fig. 1. Illustration of the periodic mesoporous organosilica DVMeP-PMO illustrating the hexagonal arrangement of monodisperse cylindrical pores (D=3.5nm) with a periodic repetition (c = 1.18nm) of silica units (orange) and ionically charged divinyl-methyl pyridinium units (yellow) along the pore axis.*

A detailed description of the synthesis methods, as well as the characterization methods that include powder X-ray diffraction and nitrogen physisorption experiments, is provided in supplementary information (Fig. S1 & S2). Remarkably, the Bragg reflections observed by X-ray diffraction have no significant counterpart in the neutron scattering intensity, which is dominated by incoherent scattering of hydrogen atoms (Fig. S3), except for the most intense (100) Bragg peak that is observable at the lowest Q-position of the QENS instrument. This is an asset for QENS experiments to separate the water and matrix contributions to the total scattering intensity of water-filled PMOs.

The structural parameters of the matrices are summarized in Table I. Although differing in ionic charges, the two bridging units have similar geometry, as confirmed by the same value of the repetition distance (1.18 nm) along the pore axis. Small differences in the porous geometry are reflected by a reduction of the pore volume, surface, and pore diameter by about 10% from 3.8 nm to 3.5 nm when comparing DVMeP-PMO to DVP-PMO. This is attributed to the presence of the additional methyl group and the counter-ion.

**Table I.** Structural parameters of the neutral (DVP-PMO) and charged (DVMeP-PMO) mesoporous matrices.

| Name | Bridging unit name | Bridging unit | Periodic distance along the pore axis (nm) [a] | Pore volume ($cm^3 \cdot g^{-1}$) [b] | Specific Surface Area ($m^2 \cdot g^{-1}$) [b] | Pore diameter (nm) |
|---|---|---|---|---|---|---|
| **DVP-PMO** | Divinyl-pyridine | (structure) | 1.18 | 0.748 | 740 | 3.8 |
| **DVMeP-PMO** | Divinyl-methyl pyridinium | (structure) | 1.18 | 0.518 | 704 | 3.5 |

[a]: evaluated from the (00l) Bragg reflections (see Supp. Info).

[b]: evaluated from the nitrogen physisorption isotherms (see Supp. Info).

All matrices were first dried under vacuum at 120°C for a minimum of 24 hours. For the preparation of hydrated matrices, a consistent amount of mesoporous material was placed in a flat aluminum rectangular cell (1 mm thick) and positioned in a desiccator alongside a beaker containing a saturated aqueous solution of NaCl or $MgCl_2$. The resulting relative humidities (RH) were 33% for the $MgCl_2$ and 75% for the NaCl at 25°C. The 75% RH exceeded the partial pressure for capillary condensation, ensuring the complete filling of the pore system, while the 33% RH is below the onset of capillary condensation, resulting in water being adsorbed onto the pore surfaces without filling the pore centers (partially filled condition). The samples were maintained in these constant humidity environments for 24 hours to ensure equilibrium was reached. Afterward, the cells were sealed with indium wire to prevent water loss and maintain a stable hydration level throughout the neutron scattering experiments.

## 2.2. Quasi-elastic neutron scattering (QENS) measurements

Quasi-elastic neutron scattering (QENS) experiments were performed on two different spectrometers, SHARP and IN16B at the Institut Laue-Langevin (ILL, Grenoble, France), to be able to probe a wide range of timescales, from ps to ns. Measurements were carried out on two different PMO powders with comparable pore geometry but different surface chemistry.

QENS methods allow measuring the incoherent neutron scattering intensity $I(Q, \omega, T)$ as a function of momentum transfer $Q$, energy transfer $E = \hbar\omega$, and temperature $T$. The accessible length scale determined by $Q$ is typically in the nanometer range. The accessible time scale is bounded between two limits determined by the energy resolution of the instrument $\delta E$ and by its dynamical range $\Delta E$. The latter two parameters can vary by orders of magnitude depending on the instrument. Finally, it should be noted that two complementary types of acquisition can be used: either a series of QENS spectra $I(Q, \omega)$ are acquired at fixed temperatures $T$, or the intensity is acquired at a fixed energy $\hbar\omega$ while the temperature is continuously ramped. The second approach can be applied either at zero energy transfer and denoted as elastic fixed window scan (EFWS), or at a fixed energy offset $\hbar\omega_{offset}$ and denoted as inelastic fixed window scan (IFWS). Bearing in mind that the instrumental energy resolution $\delta E$ is finite, EFWS does not measure purely elastic scattering but rather the fraction of scattered intensity that appears at an energy lower than $\delta E$, i.e. $I\left(Q, \omega < (\frac{\delta E}{\hbar}), T\right)$. As such, it normally increases as the dynamics slow down on cooling.

Conversely, for IFWS, $I(Q, \omega_{offset}, T)$ should exhibit a maximum around the temperature at which the typical time of the sample dynamics coincides with the time scale selected by the energy offset.

The cold neutron time-focusing time-of-flight spectrometer SHARP was used with an incident wavelength of 5.12 Å, providing an energy resolution ($\delta E$) at the elastic peak of approximately 70 µeV (FWHM), corresponding to a typical timescale ($t = \hbar/\delta E$) of about 10 ps and covering a $Q$-range from 0.2 Å$^{-1}$ to 2.0 Å$^{-1}$. It is ideally suited to measure the water fast translational and local (e.g., rotational) dynamics.

Complementing SHARP's setup, the neutron backscattering spectrometer IN16B at ILL provided a higher resolution (by about a factor of 100) suitable for slower translational dynamics. IN16B's backscattering configuration, using an unpolished Si(111) monochromator and analyzers, was set to an incident wavelength of 6.27 Å, giving an energy resolution of $\delta E$ =0.75 µeV and a timescale of around 1 ns. IN16B's measurements covered an energy transfer range $\Delta E$ of ±30 µeV and a $Q$-range between 0.19 Å$^{-1}$ and 1.84 Å$^{-1}$. IN16B's background chopper was operated in high signal-to-noise mode to enhance data quality.[35]

In both cases, temperature control was achieved with an ILL cryofurnace. Measurements on water-filled PMOs were collected sequentially on cooling to different target temperatures: 300 K, 280 K, 260 K, and 245 K on SHARP and 278 K, 258 K, and 243 K on IN16B. On SHARP, both the ionically charged DVMeP-PMO and neutral DVP-PMO were studied after their porosity was completely filled with water at 75% RH. On IN16B, the study was focused on the ionically charged matrix DVMeP-PMO, at 33% and 75% RH, allowing a specific investigation of the effect of the filling fraction. Moreover, for water-filled DVMeP-PMO (75% RH), fixed window scans (FWS) were acquired in the temperature range 195-300 K on the elastic line with Doppler drive at rest (EFWS) and at the inelastic energy offset value of 3 µeV (IFWS).[36] EFWS and IFWS probe distinct dynamical regimes: EFWS quantifies the immobile fraction of water that corresponds to energy transfer smaller than $\delta E$ (i.e., purely elastic contributions and motions with typical time exceeding one nanosecond), while IFWS detects slow motions as their inverse typical time matches with the inverse offset-energy (e.g., translational diffusion). On both instruments, the quasi-elastic spectra of the corresponding dried porous samples were additionally measured as references.

Data were processed using the MANTID software,[37] which applied corrections for detector efficiency and background contributions from the empty cell and spectrometer, to determine the scattering intensity $I(Q,\omega)$. The dynamic structure factor of confined water, $S^{water}(Q,\omega)$, was then fitted to the experimental intensity in the frequency domain using QENSH software from the Laboratoire Léon Brillouin (LLB, Saclay, France), enabling detailed interpretation of the dynamic behavior under varied temperature and humidity conditions.

## 3. Results and discussion

### 3.1. Effect of the surface charge on the water dynamics - Neutron time-of-flight study

We first consider the influence of the chemical nature of the PMO's pore surface (i.e., neutral vs charged) on the dynamics of water at full loading (i.e., 75% RH). Fig. 2 illustrates the temperature dependence of neutron scattering intensity of the two different water-filled PMOs at a representative momentum transfer $Q = 1.5$ Å$^{-1}$ for a selected range of energy transfer from -2 to +1.3 meV. From a qualitative point of view, as the temperature decreases from 300 K to 245 K, the quasi-elastic line continuously sharpens. Since the width of quasi-elastic scattering is inversely related to the timescale of particle motion, this sharpening indicates that water dynamics slow down with cooling. Also, the absence of a sharp increment in the elastic signal confirms that the water remained in a liquid state across the studied temperature range. By comparing the two panels of Fig. 2, it is evident that the water inside the ionically charged DVMeP-PMO has a sharper quasi-elastic line than the neutral DVP-PMO, indicating slower water dynamics.

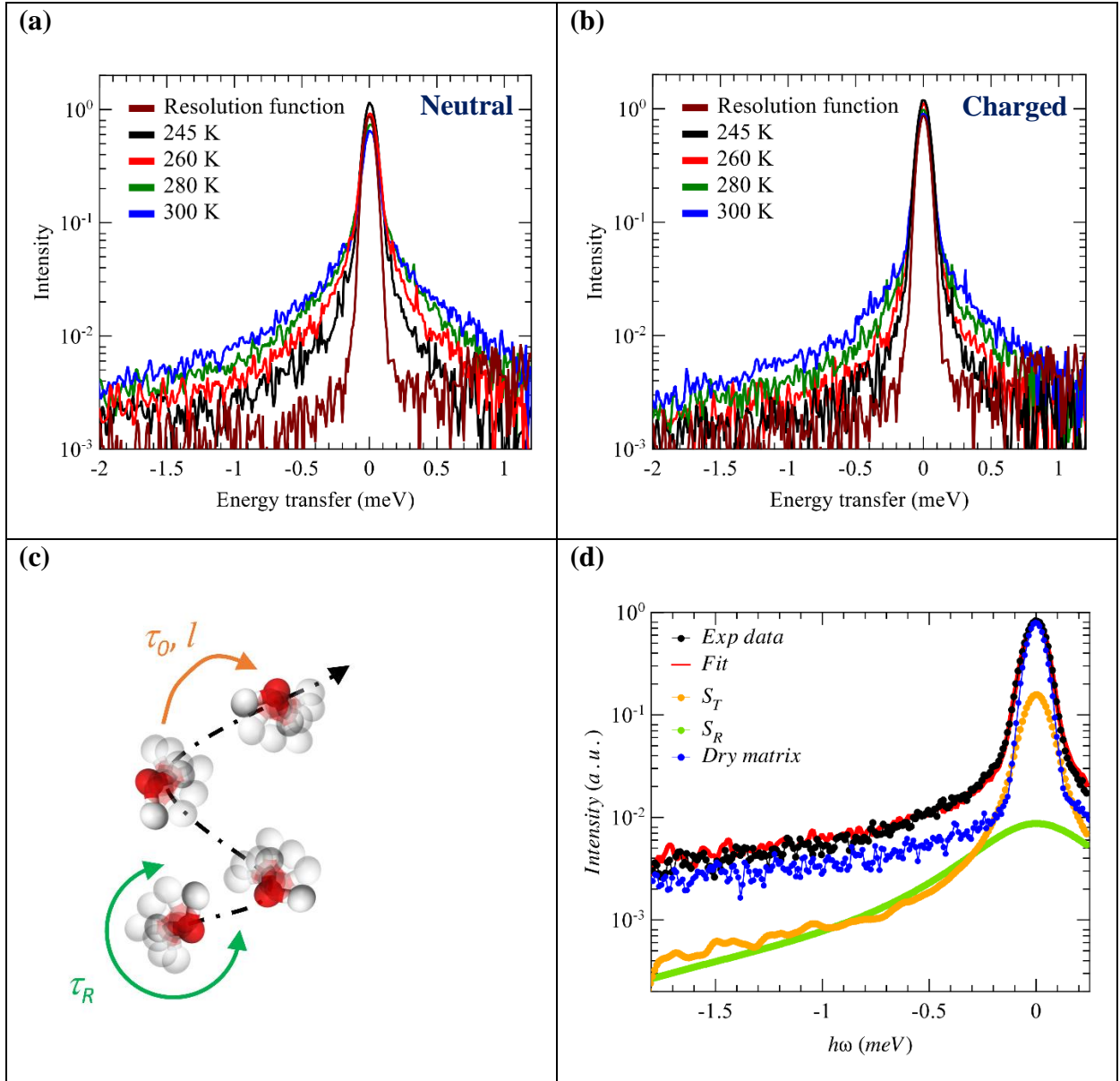

*Fig. 2. QENS spectra measured on SHARP at Q = 1.5 Å$^{-1}$ illustrate the temperature-dependent scattering intensity of water-filled (a) neutral DVP-PMO and (b) charged DVMeP-PMO. (c) Illustration of the water dynamics probed by QENS and modelled by the combination of a fast local rotation with associated correlation time $\tau_R$ and a slower jump translation diffusion with associated residence time $\tau_0$ and jump-length l. (d) Representative experimental (black) and fitted (red) scattering intensity I(Q,ω) curves at T = 280 K and Q = 1.5 Å$^{-1}$, modeled using two Lorentzian functions added to the dry matrices, for water-filled charged DVMeP-PMO acquired by SHARP.*

The scattering intensity of the water-filled matrix, $I^{water\,filled}(Q,\omega)$, was fitted at each individual $Q$ by a model that was the sum of the scattering intensity from the matrix, corresponding to the experimental intensity of the dry matrix $I^{dry}(Q,\omega)$, and the scattering from the confined water convoluted by the instrumental resolution function $R(Q,\omega)$. A detailed description of the underlying theoretical model and the fitting procedure is provided as supplementary information.

In agreement with the approach of recent studies on confined water in MCM-41 and PMOs with different surface functionality,[25] we found that the dynamic structure factor of water $S^{water}(Q,\omega)$ could be modelled by the combination of two distinct motions: a slow translational diffusion ($T$) and a faster local rotation-like motion ($R$). We defined $S_T(Q,\omega)$ and $S_R(Q,\omega)$ the dynamic structure factors of these motions. The diffusive motion was modelled by a Lorentzian quasielastic term

$$S_T(Q,\omega) = L_T(Q,\omega) \qquad (1)$$

while the second motion comprised both an elastic and an inelastic term due to its localized character

$$S_R(Q,\omega) = A(Q)\,\delta(\omega) + \big(1 - A(Q)\big)L_R(Q,\omega) \qquad (2)$$

The spectral widths of these two motions (HWHM) are noted $\Gamma_T(Q)$ and $\Gamma_R(Q)$, respectively, and $A(Q)$ is the elastic incoherent structure factor (EISF) of the local dynamics.

The relative contribution of the broadest component ($L_R$) to $S^{water}(Q,\omega)$ was found to be negligible at low-$Q$, and even at room temperature, it could be determined accurately only for values of the transfer of momentum larger than $Q = 1.2$ Å$^{-1}$. This is fully consistent with the localized nature of the associated motion, leading to a large value of the corresponding EISF $A(Q)$, which eventually approaches unity at small values of $Q$. In this situation, the intensity of the quasielastic component (second term of Eq. 2) vanishes, and only the elastic term persists. The EISF is directly related to the spatial trajectory of the motion. In the classical model of water proposed by Teixeira et al., the local hydrogen dynamics are described as isotropic rotations around the oxygen atom. The associated EISF is given by $A_R(Q) = [j_0(QR_R)]^2$, where $j_0$ denotes the zeroth-order spherical Bessel function and $R_R = 0.98$ Å represents the length of the O–H bond.[38]

For water confined in neutral and charged PMOs, this classical model only showed good agreement with experimental data at high temperature (300 K) and in the high $Q$ range (see Fig. S5). By decreasing the temperature, the measured EISFs were consistently higher than this theoretical model, indicating that hydrogen trajectories are more spatially restricted than expected for free rotational motions. This means that the broad quasi-elastic line cannot be solely attributed to isotropic rotational diffusion, but rather interpreted in terms of restricted rotation or librational motion. The associated correlation time, as determined by the inverse half-width at half-maximum (HWHM) of the broad Lorentzian component ($\Gamma_R$), is on the order of 1-3 ps (Fig. S7), which is in line with previous studies on confined water.[25] Compared to DVP-PMO, the local dynamics of water filled in charged DVMeP-PMO appeared slower ($\Gamma_R$ smaller by about a factor of 1.5 to 2) and spatially more restricted with larger EISF as illustrated in Fig. S6. This can be related to the strengthening of the water surface interaction through H-bonds and solvation in the ionically charged PMO compared to its neutral counterpart.

We now consider the most significant component of the dynamic structure factor in the studied range of temperature, energy, and transfer of momentum, which is the first Lorentzian term, $L_T(Q,\omega)$ attributed to translation dynamics, as sketched in Fig. 3(a) and 3(b). The temperature dependence of its HWHM ($\Gamma_T$) as a function of $Q^2$ is illustrated in Fig. 3(c) and 3(d). A direct visual comparison of the two panels obviously demonstrates that the water dynamics in the charged DVMeP-PMO are significantly reduced compared to the water-filled neutral DVP-PMO. This striking slowdown induced by different surface chemistry is accompanied by a systematic reduction in mobility with decreasing temperature.

For most of the data, $\Gamma_T(Q^2)$ exhibits an initial linear evolution at low $Q^2$, followed by a deviation at higher $Q^2$ values upon reaching a plateau. In fact, the initial linear increase of $\Gamma_T(Q^2)$ with $Q^2$ reflects diffusive motion, which is in agreement with Fickian behavior and also consistent with the absence of an elastic term that would have indicated that diffusion was spatially restricted on these specific time and length scales. Conversely, in the high $Q$ accessible range, $\Gamma_T(Q^2)$ deviated from the Fickian behavior and reached a plateau. This indicates that translation motions are discontinuous at the molecular scale. This plateau is more pronounced in the charged PMO due to longer residence times, as expected from strong water-surface interactions. This behavior is consistent with previous findings regarding the water dynamics in nanoconfined media.[25, 26, 38] It is well-described by a jump-diffusion model, suggesting that the translational motion of water molecules is intermittently interrupted by localized residence times. To quantitatively describe this behavior, we fitted the data using the Singwi and Sjölander (SS) model, which assumes an exponential distribution of jump lengths.[39] According to this model, the $Q^2$-dependence of $\Gamma_T$ is given by:

$$\Gamma_T = \frac{D_T Q^2}{1 + D_T Q^2 \tau_0} \qquad (4)$$

where $\tau_0$ is the residence time between jumps, and $D_T$ is the translational diffusion coefficient.

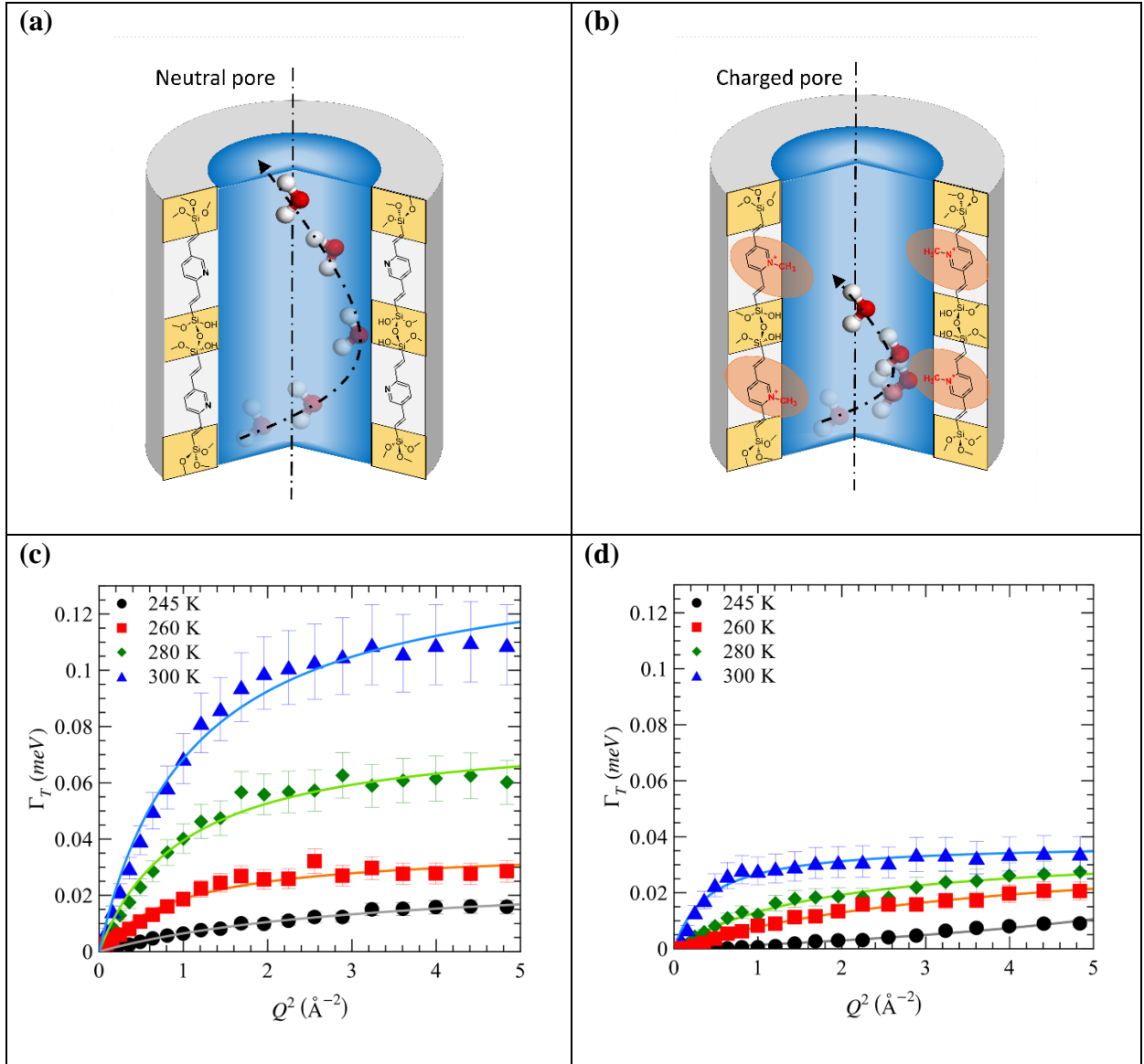

*Fig. 3. Illustration of the trajectory of water in (a) neutral DVP-PMO and (b) charged DVMeP-PMO, indicating the much longer residence time in the latter case, resulting in a significant slowdown of translational diffusion. Evolution of the half-width at half-maximum of the sharp Lorentzian ($\Gamma_T$) as a function of $Q^2$, obtained from the fitting of QENS spectra measured on SHARP for water confined in (c) neutral DVP-PMO and (d) charged DVMeP-PMO at four different temperatures. The fits using the jump-diffusion model are shown as solid lines.*

The fitting parameters are summarized in Table II. As shown in Fig. 4(a), the results in their Arrhenius plot indicate that across all temperatures, the diffusion coefficient $D_T$ for water confined within neutral PMOs is only slightly smaller than that of bulk water, especially at 300 K ($1.76 \times 10^{-9}$ m$^2$s$^{-1}$ vs. $2.3 \times 10^{-9}$ m$^2$s$^{-1}$). This moderated effect of confinement aligns with previous observations made for another non-ionic PMO, denoted DVB-PMO in Fig. 4(a), which presents a pore geometry comparable to DVP-PMO but a more hydrophobic character conveyed by divinylbenzene bridging units.[25] Moreover, Fig. 4(b) shows that the residence time $\tau_0$ is prolonged for water in neutral DVP-PMO compared to bulk water at the same temperature (7 ps vs. 1.1 ps at 300 K for bulk water). This increase in residence time is slightly larger than previously reported for water confined in DVB-PMO. This difference is compatible with the existence of interfacial H-bonds between water and the pyridine group for DVP-PMO, which is absent for DVB-PMO.

From this viewpoint, the situation encountered for divinyl-pyridine shares a similarity with that reported for the hydrophilic divinyl-aniline bridged PMO (DVA-PMO) by Jani et al..[25] In the latter case, it was found that the amino groups of the bridging units acted as secondary H-bonding sites for interfacial water molecules in addition to silanols of the silica units. This was supported by the NMR observation of strong interfacial correlations between water and both silica and organic units.[28] Also, neutral PMOs prepared using divinyl-pyridine precursors possess a weakly basic surface due to the protonation equilibrium of pyridine groups (pKa ≈ 5.2 for pyridine). The partial protonation induces mild electrostatic interactions that have minimal impact on water diffusion. Therefore, water confined in neutral DVP-PMO retains near-bulk translational diffusion, although slightly reduced compared to other neutral PMO with a higher hydrophobic character, such as DVB-PMO.

A starkly different behavior is observed for water within the ionically charged DVMeP-PMO, where the diffusion coefficient is significantly reduced, approximately by a factor of four (i.e., $D_T$ = $0.6 \times 10^{-9}$ m$^2$s$^{-1}$ at 300 K). More strikingly, the residence time is one order of magnitude longer for water confined in charged DVMeP-PMO (i.e., 20.5 ps at 300 K) than for bulk water. This slowdown is considerably larger than the confinement effects observed in the present study for neutral DVP-PMO and for many other water-filled PMOs and silicas with comparable pore size, whatever the hydrophilic character of their surface.[25] It is noteworthy that the 8% reduction in pore size between DVP-PMO (3.8nm) and DVMeP-PMO (3.5 nm) cannot be considered as a significant reason for the observed slowdown. Indeed, in the QENS study on water confined in two neutral PMO that presented a difference in pore size of 15%, namely DVB-PMO (4.1 nm) and BP-PMO (3.5nm), it was observed only a minor reduction of 10% in the diffusivity and an increase of 50% in the residence time.[25] Therefore. we can conclude that the ionic character of the DVMeP-PMO compared to neutral DVP-PMO actually plays the crucial role in the observed slowdown.

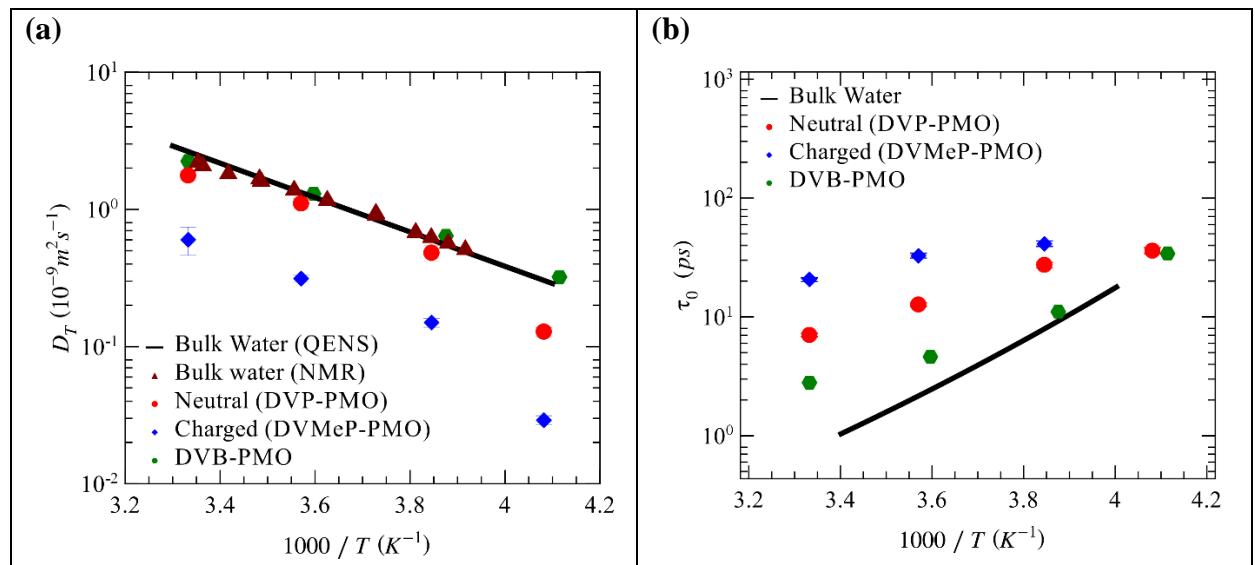

*Fig. 4. (a) Translational diffusion coefficient ($D_T$) of water and (b) residence time ($\tau_0$) of water, evaluated from the fit of SHARP spectra using the jump-diffusion model, for water confined in neutral (red circle) and charged (blue diamond) PMOs as a function of inverse temperature. For comparison, QENS data for bulk water from Teixeira et al.[38] and NMR diffusion coefficients from Price et al.[40] are presented alongside QENS data for DVB-PMO from Jani et al..[25] Adapted with permission from [Teixeira et al., 1985]. Copyright [1985] American Physical Society. Adapted from [Price et al., 1999]. Copyright [1999] American Chemical Society.*

The combination of the diffusion coefficient and the residence time can be used to evaluate the mean squared jump length $\langle l^2 \rangle$ that is expressed as:

$$\langle l^2 \rangle = 6 D_T \tau_0 \quad (5)$$

The jump length values are in the range of 2 to 3 Å for both PMOs, with no systematic temperature dependence. This is about twice as large as for bulk water, which suggests that the microscopic mechanism involved in jump diffusion is altered in confined geometry, possibly as a result of a disrupted H-bond network. The most striking observation is that the jump length has comparable values for the neutral and ionically charged PMO, despite very different values of the diffusion coefficient and the residence time. This scale is far smaller than the pore size (~3.5–3.8 nm), meaning that the observed motion reflects local rearrangements within a structured hydrogen-bond network, not long-range diffusion limited by the pore walls. In this picture, the jump itself likely corresponds to a displacement of a water molecule from one local site (i.e., basin) of attraction (hydrogen-bonded configuration) to another. It suggests that the topology of these local sites does not change drastically between the neutral and charged environments. What differs significantly, however, is the residence time within each site (i.e., how long a water molecule remains in a given local configuration before making a jump). In the charged PMO, electrostatic interactions with the pyridinium groups deepen these sites or raise the energy barriers (i.e., activation energy) between them, leading to longer residence times and therefore a reduced diffusion coefficient. In contrast, the neutral PMO allows for more frequent transitions between sites, resulting in faster diffusion.

**Table II.** *Fit parameters $D_T$, $\tau_0$, and $l$, obtained by fitting of $\Gamma_T(Q^2)$ using the jump-diffusion model for data acquired by SHARP.*

| | Neutral | | | Charged | | | Bulk | | |
|---|---|---|---|---|---|---|---|---|---|
| T (K) | $D_T$ ($10^{-9}$ m$^2$/s) | $\tau_0$ (ps) | $l$ (Å) | $D_T$ ($10^{-9}$ m$^2$/s) | $\tau_0$ (ps) | $l$ (Å) | $D_T$ ($10^{-9}$ m$^2$/s) | $\tau_0$ (ps) | $l$ (Å) |
| 245 | 0.13 ± 0.01 | 36 ± 2 | 1.7 | 0.03 ± 0.002 | - | - | 0.42 | 22.7 | 2.4 |
| 260 | 0.49 ± 0.06 | 27 ± 1.5 | 2.8 | 0.15 ± 0.01 | 41 ± 2 | 1.9 | 0.56 | 8.9 | 1.7 |
| 280 | 1.11 ± 0.09 | 13 ± 0.5 | 2.9 | 0.31 ± 0.02 | 32.5 ± 1.5 | 2.5 | 1.3 | 2.3 | 1.3 |
| 300 | 1.76 ± 0.13 | 7.0 ± 0.3 | 2.7 | 0.6 ± 0.15 | 20.5 ± 1 | 2.7 | 2.3 | 1.1 | 1.2 |

In general, the jump-diffusion model predicts that the low-range diffusion coefficient is determined at the microscopic scale both by the typical length between two successive atomic jumps and the residence time that separates these two events. Within this theoretical framework, it can be deduced that the reduction in the translational diffusion of water within charged DVMeP-PMO is primarily due to the huge increase in the residence time. This is indicative of the enhanced electrostatic interactions between the water molecules and the pyridinium groups, and the counterion present on the surface of the DVMeP-PMO. On average, the surface charge is balanced by the counterions. However, at the molecular scale, one expects that the dissociation of the pyridinium and iodide ion pairs is enabled by the presence of water. As a result, highly polar and positively charged surface areas are formed around pyridinium groups. They act as strong surface adsorbing sites that strengthen hydrogen bonding and alter water dynamics. As for the iodine counterion, it is worth noting that a chaotropic effect could also be invoked. In fact, chaotropic ions are expected to weaken the hydrogen-bond network in water and increase the water dynamics.[26] Although the iodide counterion is chaotropic and expected to promote increased water mobility, our results show a substantial reduction in translational diffusion. This suggests that the dominant factor is actually the strong electrostatic interaction between water and the pyridinium group present at the pore surface, which overcomes the chaotropic effect of iodine counterion on the water dynamics. Indeed, the existence of ionically charged adsorption sites and the resulting strong electrostatic interactions can favor long-lived interfacial environments that differ from the hydrogen-bond network of water found in normal conditions. Being more tightly bound to the charged surface, the water molecules exhibit longer residence time and, therefore, reduced translational mobility.

Whether this conclusion can be generalized to other systems is an open question, which highlights the interest in investing further effort in the design of PMO with different chaotropic or kosmotropic counterions. Indeed, a different situation may be encountered for other counterions, resulting from a different balance between the respective effects of water-pore and water-counterion interaction on water dynamics. The impact of counterion identity, particularly the hydration strength and polarizability, is an important variable that remains to be systematically addressed. While I⁻ was used in this study due to its synthetic accessibility via methyl iodide, future studies would explore Cl⁻, Br⁻, and OH⁻ to determine how counterions modulate confined water dynamics.

To place our findings in a broader framework on how confinement and surface charge influence phase behavior, it is also worth mentioning previous X-ray diffraction studies that investigated the effects of pore surface chemistry, and more specifically surface ionic charge, on confined ice and water structures in PMOs.[27, 41] Importantly, those studies observed ice formation only at 230 K, which is fully consistent with our current observation that the dynamics measured by QENS at 243 K and above are characteristic of the liquid state.

Despite the different temperature range and so, the different nature of the confined phase, this work also indicates that the structural behavior of confined water is shaped by the combined effects of surface chemistry and geometric confinement. Prior work has shown that in larger pores (e.g., $d_p$ = 4.9 nm), stacking-disordered ice (Isd), comprising both hexagonal and cubic phases, can form under specific conditions. In contrast, the smaller pore sizes used here ($d_p$ = 3.5–3.8 nm) impose stronger confinement that suppresses crystallization.[41] For charged PMOs, the high polarity and hydrogen-bonding capacity of pyridinium groups induce local structuring of confined water and deviations in lattice order, even in the absence of ice formation. These effects are amplified by the smaller pore size in DVMeP-PMO ($d_p$ = 3.5 nm). In neutral PMOs ($d_p$ = 3.8 nm), weaker electrostatic interactions allow more bulk-like water behavior in the pore center, while the partially protonated pyridine groups still generate interfacial structuring near the walls. These observations underscore the role of surface functionalization, not just in modifying interfacial behavior but also in tuning phase stability and dynamic regimes throughout the pore.

Moreover, a recent study investigating the OH-stretching band of water near graphene with varying surface charges revealed that water molecules at positively charged interfaces, similar to those in charged PMOs, exhibit significantly stronger hydrogen bonding.[42] This enhanced hydrogen-bonding network causes water molecules to bind more tightly to the solid surface, resulting in reduced mobility and increased interfacial structuring.

Our findings not only support the conclusion about the critical role of surface chemistry that was derived in previous studies of water-filled non-ionic PMOs with variable hydrophilicity, but they also demonstrate the unique effect of ionic surface charge. Consequently, the question emerges about the spatial extension of the surface charge effect on the water dynamics. Is it strictly limited to those water molecules that interact directly with the surface ions through adsorption, H-bonding, and solvation forces, or does it spread over longer distances in the confined liquid? To address this issue, complementary QENS experiments with variable water loading were performed on DVMeP-PMO.

## 3.2. Effect of the filling fraction on the water dynamics - Neutron backscattering study

To explore water behavior near the charged surface in more detail, we studied the effect of varying the amount of adsorbed water in DVMeP-PMO by tuning the relative humidity. From the previous part, it is clear that the huge slowdown of the dynamics of water in charged PMOs makes such a systematic study, if possible, extremely challenging on SHARP. The energy resolution of the IN16B neutron backscattering spectrometer is about two orders of magnitude better than that of SHARP, but its dynamical range is also limited. This makes it unsuitable to study the water dynamics in the neutral PMO at the same temperatures, but it is ideally suited to focus on the case of ionically charged DVMeP-PMO. The selected values, 75% and 33% RH, are located respectively above and below the capillary condensation pressure. They made it possible to compare the dynamics of the water in the completely filled pore at 75% RH to the water adsorbed approximately two to three molecular layers on the surface of the pore at 33% RH, as illustrated in Fig. 5(a) and Fig. 5(b), respectively.

Fig. 5(c) and Fig. 5(d) display the temperature-dependent scattering intensity of IN16B spectra at a momentum transfer $Q = 1.5$ Å$^{-1}$ over an energy transfer range of -30 to +30 μeV for fully-filled and partially-filled charged PMO, respectively. As the temperature decreases from 278 K to 243 K, the quasi-elastic line continuously sharpens, a trend previously observed in the SHARP data. This sharpening reflects the slowing down of water dynamics at lower temperatures, as the narrowing of the quasi-elastic width indicates longer timescales of molecular motion. This interpretation is fully supported by the elastic fixed window scan (EFWS) measured at 75% RH (Fig. S9) that continuously increases on cooling from 280 K to 190 K, with a sigmoidal shape centered at about 230 K. In the meantime, the inelastic fixed window scan (IFWS) measured at 3 μeV presents a single peak that covers the temperature range from 220 K to 265 K at half maximum intensity, and that is centered around 235 K. Its maximum position shifts to a higher temperature as the value of the transfer of momentum decreases as illustrated in Fig. 5(e). This evolution reveals the dispersive nature of the associated dynamics, which is attributed to the Q-dependent onset of translational diffusion dynamics within the dynamical range fixed by the selected off-set energy (3 μeV). It also demonstrates that below 200 K, the subsequent increase of elastic intensity is not related to translational dynamics anymore, but rather to localized modes (e.g., vibrations) that contribute to the Debye-Waller factor.

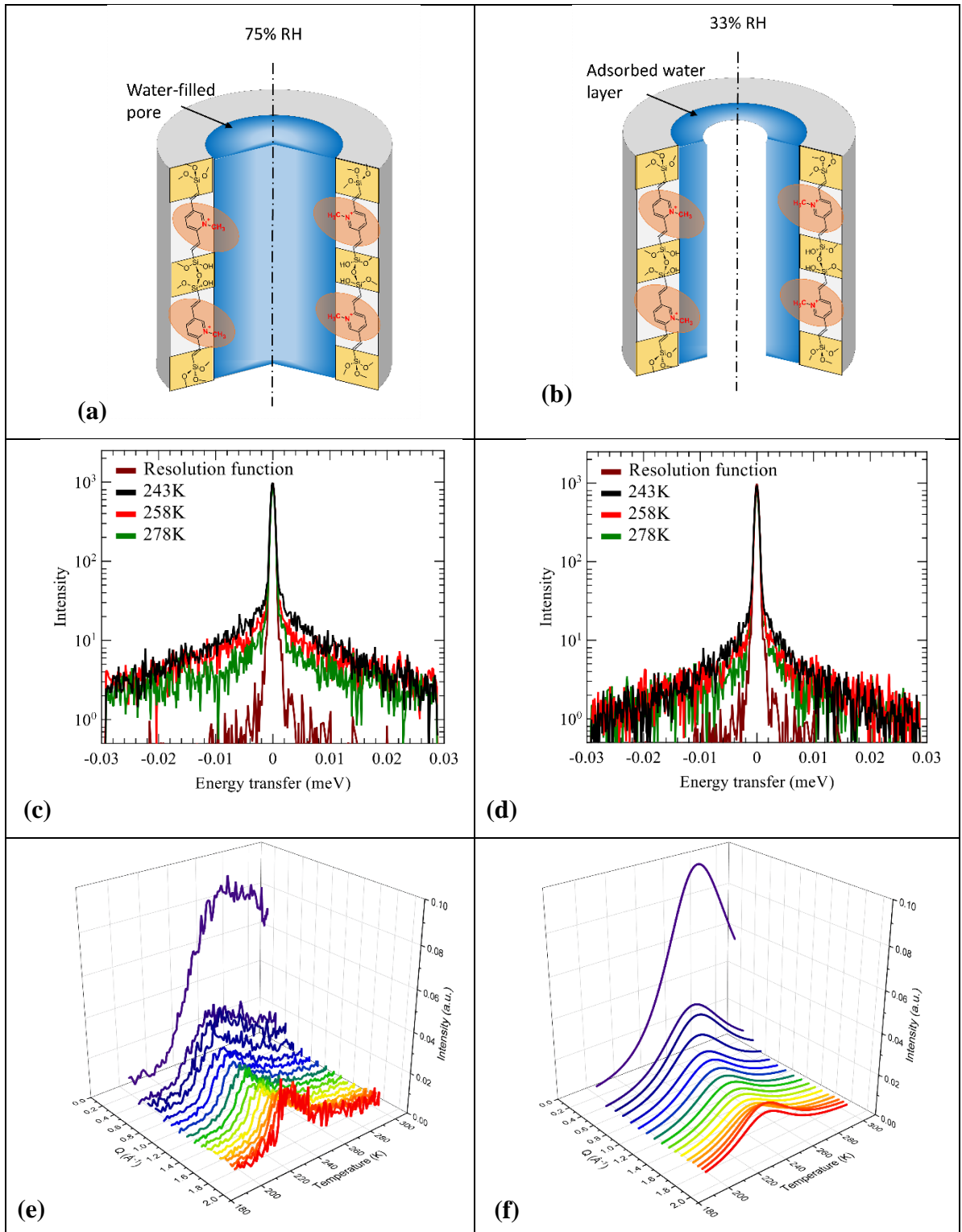

*Fig. 5. Illustration of the (a) fully-filled (75% RH) and (b) partially-filled (33% RH) charged DVMeP-PMO. QENS spectra of the (c) fully-filled and (d) partially-filled charged DVMeP-PMO measured on IN16B at $Q = 1.5 \text{ Å}^{-1}$ and three different temperatures. Inelastic Fixed Window Scans (IFWS) for fully-filled, charged DVMeP-PMO as a function of temperature and transfer of momentum for $Q \geq 0.3 \text{Å}^{-1}$. The experimental intensity (e) and to corresponding fitted intensity (f).*

To quantitatively analyze the spectra, each spectrum was individually fitted at each $Q$-value using a model that incorporated a single Lorentzian function and a flat background (Fig. S5). The procedure of fit and the theoretical model expressed in Eq. 6 are similar to those used for SHARP. The only exception concerns the second Lorentzian contribution $L_R(Q,\omega)$ that was related to the quasi-elastic component of the local rotational dynamics. According to SHARP study, the width of $L_R(Q,\omega)$ considerably exceeds the dynamical window of IN16B. Therefore, it was replaced by a flat $\omega$-independent background $BG(Q)$.

The fitted incoherent scattering intensity can then be expressed as:

$$I_{QENS}^{fit}(Q,\omega) = I^{dry}(Q,\omega) + R(Q,\omega) \otimes [A(Q)L_T(Q,\omega) + BG(Q)] \quad (6)$$

As a result, only the slower component related to translational dynamics was evaluated on IN16B. Based on the same hypothesis, the IFWS was fitted by Eq. 7

$$I_{IFWS}^{fit}(Q,T) = A(Q)L_T(Q,\omega_{offset}) + BG(Q) \quad (7)$$

where $\hbar\omega_{offset}$ (= 3 µeV) represents the offset energy, as illustrated in Fig. 5(f). Here, the temperature dependence of the linewidth $\Gamma(Q)$ of the Lorentzian function $L_T$ was assigned to the Arrhenius law, yielding an activation energy $Ea$ = 28.1 kJ.mol$^{-1}$ as expressed by Eq. 8. This value is in excellent agreement with the activation energy obtained from $I_{QENS}^{fit}$ and Fig. 7(a) (Ea = 28.47 kJ mol$^{-1}$), demonstrating the consistency of both methods (QENS and IFWS) in describing the dynamical process.

$$\Gamma_{IFWS}(Q,T) = \Gamma_0(Q)exp\left(\frac{-E_a}{k_BT}\right) \quad (8)$$

The linewidth (HWHM) of the quasi-elastic intensity measured on IN16B is shown in Fig. 6 at three temperatures, 243 K, 258 K, and 278 K. It initially follows a trend proportional to the square of the momentum transfer ($Q^2$) but gradually deviates from the linear slope at higher $Q$-values, as observed on SHARP. In certain cases, particularly at higher temperatures, the linewidth displays fluctuations rather than a smooth, monotonic increase. These fluctuations are attributed to statistical noise, more prominent at higher temperatures when the quasi-elastic broadening approaches the maximum energy transfer of 30 µeV on IN16B. At lower temperatures, the quasi-elastic intensity increases, and these fluctuations diminish as the linewidth sharpens.

From the fitting parameters of IFWS, the linewidths $\Gamma_{IFWS}(Q,T)$ were independently evaluated at 243 K and 258 K, as indicated by open symbols in Fig. 6(a). They showed excellent agreement with those evaluated from QENS spectra. Such a comparison was elusive at 278 K, because the quasi-elastic broadening systematically exceeds the selected energy offset indicated by the arrow in Fig. 6(a). This is also illustrated in Figure S9, which confirms that only the two lowest temperatures fall within the temperature range of the IFWS peak.

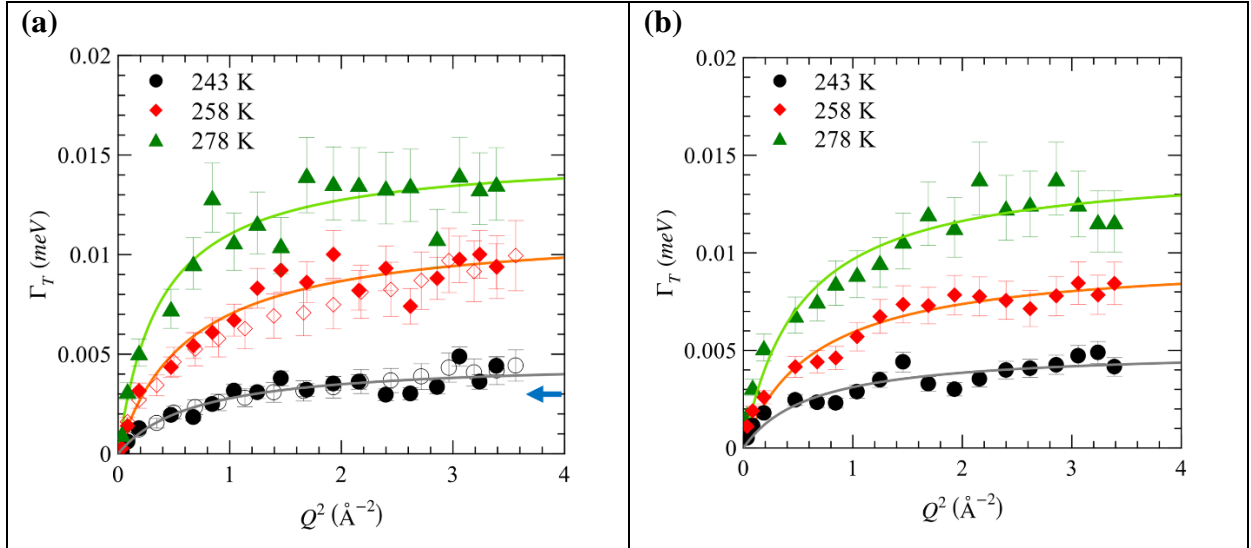

*Fig. 6. Evolution of the half-width at half-maximum of Lorentzian ($\Gamma_T$) as a function of $Q^2$, obtained from the fitting of QENS spectra measured on IN16B for water confined in (a) fully-filled and (b) partially-filled charged PMOs at three different temperatures. The fits using the jump-diffusion model are shown as solid lines. The open symbols represent the $\Gamma_T$ value obtained through an inelastic fixed window scan (IFWS). The blue arrow in the left panel shows the energy off-set of 3 µeV that was used for IFWS.*

As for SHARP measurements, the linewidth data were fitted using a jump-diffusion model (Eq. 4). The resulting diffusion coefficients and residence times, presented in Fig. 7 and summarized in Table III. It should be noted that the values of the diffusion coefficient and residence times obtained on IN16B for the fully-filled DVMeP-PMO systematically differ (by a factor of 1.8-2.5) from those measured on SHARP. This stems from the vastly different energy resolutions of the two spectrometers. SHARP, with a coarser resolution ($\delta E \approx 70$ µeV), captures faster dynamics and is thus sensitive to broader quasi-elastic features. IN16B, with a two order of magnitude finer resolution ($\delta E \approx 0.75$ µeV), probes significantly slower motion and captures only the narrowest components of the dynamics. These differences reflect the timescale sensitivity of the instruments, with different limitations and not an inconsistency in the underlying physics. The impact of the instrument resolution is especially significant for water in charged DVMeP-PMO that exhibits a drastic slowdown compared to bulk water. Indeed, its quasielastic broadening (~10 µeV at 280K and $Q = 1$ Å$^{-1}$) entered the elastic resolution of SHARP, while it was obviously not the case on IN16B. Note that the opposite situation was encountered for water in neutral DVP-PMO, with a quasielastic broadening (~40 µeV at 280K and $Q = 1$ Å$^{-1}$) that was ideally-suited for SHARP but would have exceeded the dynamical range of IN16B ($E_{max} = \pm 30$ µeV).

We compared in Fig. 7 the diffusion coefficient and residence time values obtained on IN16B for water in charged DVMeP-PMO with those obtained on the same instrument for a water-filled non-ionic DVB-PMO by Jani et al..[25] In agreement with the observation made for SHARP data, the diffusion coefficient $D_T$ for water confined in the charged DVMeP-PMO is consistently lower than for water in the non-ionic DVB-PMO. Also, the characteristic residence times ($\tau_0$) are significantly higher in the charged matrices compared to the neutral systems. This confirms that the water molecules in charged DVMeP-PMO exhibit much slower diffusion due to stronger electrostatic interactions with the surface ionic groups.

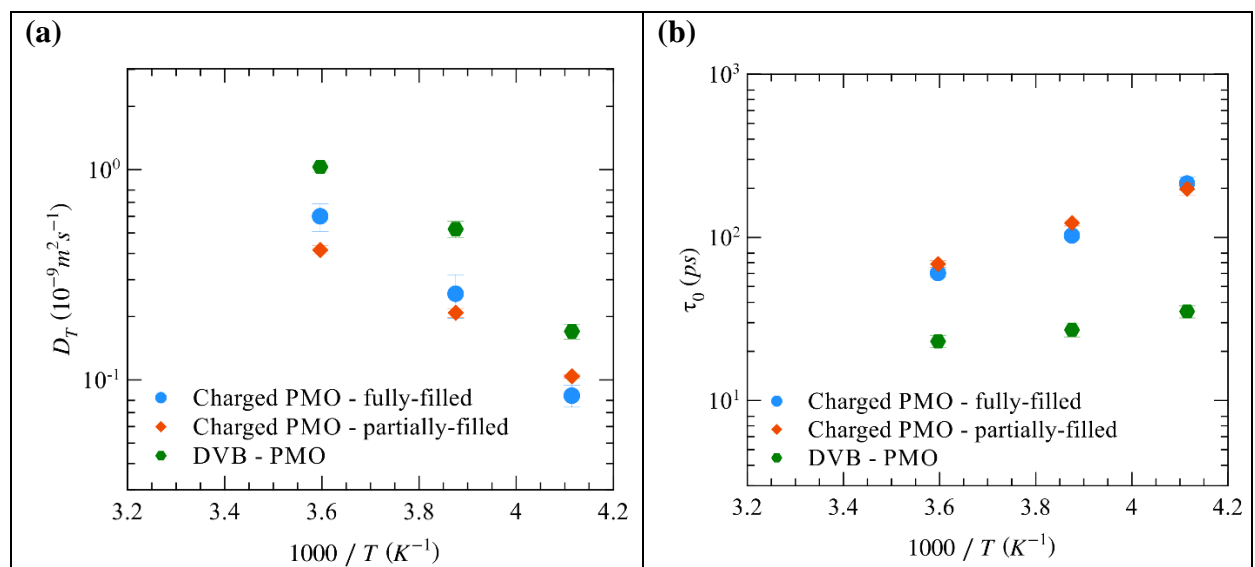

*Fig. 7. (a) Translational diffusion coefficient ($D_T$) and (b) residence time ($\tau_0$) of water, evaluated from the fitting of IN16B spectra using the jump-diffusion model, for water confined in fully-filled and partially-filled charged DVMeP-PMO as a function of inverse temperature. QENS data for $D_T$ and $\tau_0$ in DVB-PMO, as derived from Jani et al.,[25] are also included.*

The new insight provided by the experiments on IN16B concerns the effect of the filling ratio. In fact, the diffusive dynamics of water in the partially filled DVMeP-PMO (2-3 interfacial water at 33% RH) are slower than that of water in the same matrix at full loading (capillary filled water at 75% RH). However, this additional slowdown is rather modest: at 278 K, the diffusion coefficient for water adsorbed at 33% RH has only been reduced by 30% compared with water in capillary-filled pores at 75% RH. Similarly, the increase in residence time is only about 22%. This observation made for charged PMOs contrasts sharply with the approximately 4x reduction in diffusivity observed between neutral and charged PMOs at full loading. It underscores that varying the interfacial layer thickness of water adsorbed in this specific DVMeP-PMO ionically charged matrix has a much smaller impact on the water dynamics than varying the nature of the pore surface (neutral vs. charged) for a fully filled pore. Moreover, the huge slowdown effect induced by the surface charge is maintained over the entire temperature range studied, while its mild dependence on RH tends to disappear upon decreasing the temperature to 243 K, as shown in Fig. 7.

This is an important indication that the huge impact of the ionic interaction on the water dynamics is not restricted to the adsorbed molecules located at the pore surface but affects the water molecules present in the entire pore volume. This can be explained by the long-range nature of electrostatic interaction that remains significant over a distance that approaches the pore radius of 1.8 nm. The specific H-bonded structure adopted by water molecules can also play an important role in connecting the dynamics of the water molecules located in the pore center to those in direct interaction with the surface charged units. On cooling, the observed merging of water dynamics in partially and fully filled pores can also be interpreted by the H-bond network becoming stronger at low temperatures, so that the liquid volume that is influenced by the surface condition increases.

**Table III.** *Fit parameters $D_T$, $\tau_0$, and l, obtained by fitting $\Gamma_T(Q^2)$ using the jump-diffusion model for data acquired by IN16B.*

|  | **Fully-filled** | | | **Partially-filled** | | |
|---|---|---|---|---|---|---|
| **T (K)** | $D_T$ ($10^{-9}$ m$^2$/s) | $\tau_0$ (ps) | $l$ (Å) | $D_T$ ($10^{-9}$ m$^2$/s) | $\tau_0$ (ps) | $l$ (Å) |
| 243 | 0.10 ± 0.01 | 213 ± 20 | 3.6 | 0.12 ± 0.002 | 206 ± 17 | 3.9 |
| 258 | 0.27 ± 0.06 | 102 ± 5.5 | 4.1 | 0.22 ± 0.01 | 132 ± 5 | 4.1 |
| 278 | 0.61 ± 0.09 | 60 ± 3 | 4.7 | 0.43 ± 0.02 | 73 ± 4 | 4.4 |

## Conclusion

This work explored water confined in neutral and charged periodic mesoporous organosilicas (PMOs), revealing how surface chemistry (i.e., surface charge and its polarity) influences molecular dynamics. Using neutron scattering techniques (SHARP and IN16B), we demonstrated the distinct effects of divinylpyridine-based neutral PMOs and their methylated, ionically charged counterparts on water dynamics.

Comparing neutral and charged PMOs highlights the critical role of surface charge in modulating water dynamics, and the results provide detailed quantitative measurements of translational diffusion and residence times, offering insights into the crucial role of molecular interactions and electrostatic effects in modulating water mobility under confinement.

The most salient phenomenon is observed for the ionically charged PMO, which induces an extreme slowdown of the water translational dynamics. Based on the jump-diffusion modelling of the QENS spectra, this decrease in the dynamics can be quantified as a reduction by a factor 4 of the diffusion coefficient and an increase by one order of magnitude of the residence time. Such a colossal slowdown can be attributed to the strong surface adsorption of water on the cationic pyridinium sites, which induces long-lived interfacial structures. This conclusion is supported by the fact that the mean jump length barely depends on the surface charge, which means that the reduction of the diffusion coefficient is solely a consequence of the much longer residence time.

Remarkably, the impact of the surface charge is not restricted to the dynamics of the adsorbed water molecules but affects the entire volume of the pore with a pore size of 3.7 nm. This important conclusion can be derived from the high resolution QENS study of PMO hydrated at different RH, which demonstrates a modest difference between the dynamics of the fully filled sample compared to the adsorbed liquid layer.

The dynamical properties of water confined in surface charged PMOs are unusual and markedly different from what is currently known from neutral porous media. In the latter cases, water dynamics exhibit only a moderate slowdown, with the extent of the effect depending on the chemical nature of the pore surface. This phenomenon is traditionally attributed to varying degrees of hydrophobicity and hydrophilicity. In the present study, this is confirmed for water-filled DVP-PMO, which exhibits a hydrophilic character further enhanced by the partial protonation of pyridine groups. Consequently, the confinement effect on the water dynamics exceeds the observations made in the literature for other neutral PMO, such as DVB-PMO, but it remains much smaller than for the charged DVMeP-PMO.

These findings underscore the importance of tailoring surface chemistry to control water dynamics in confined systems. While this study provides valuable insights into water dynamics in neutral and charged PMOs, it also raises several questions warranting further investigation. For example, the role of different counterions and varying degrees of functionalization in modifying water dynamics within charged PMOs remains an open area of research, particularly at lower hydration levels where water-ion interactions could dominate. Finally, integrating experimental data with molecular dynamics simulations, mapping hydration structure and dynamics near pyridinium-functionalized surfaces, could provide a more detailed atomistic view of water-surface interactions and help elucidate the mechanisms underlying the observed differences in diffusion and residence times. Such insights would enhance our ability to predict and design materials with desired water transport properties.

## Conflicts of interest

There are no conflicts to declare.

## Supporting Information

PMO synthesis. Structural characterization of Divinyl Pyridine Bridged PMO and Divinyl-methyl Pyridinium Bridged PMO. Neutron structure factor of dry and capillary filled PMO. Fitting procedure. Elastic Incoherent Structure Factor (EISF) of the local quasi-elastic relaxation obtained from SHARP measurements. Timescale of rotational and librational motions. Arrhenius fitting of the translational diffusion coefficient of water in neutral and charged PMOs. Elastic and Inelastic Fixed Window Scans measured on IN16B.

## Acknowledgments

This research was carried out as part of the DFG-ANR collaborative project (ANR-23-CE29-0028) for which we express our gratitude. DM and AM acknowledge funding by ANR-22-CE50-0002. We also thank the Institut Laue-Langevin and 2FDN (Fédération Française de Diffusion Neutronique) for providing neutron beam time (IN16B #6-07-90 and SHARP #CRG-2896), which was essential to this work. Scientific exchange and financial support of the Center for Molecular

Water Science (CMWS), the research initiative "BlueMat: Water-Driven Materials" at Hamburg University of Technology, the Cluster of Excellence "Advanced Imaging of Matter" of the DFG (EXC 2056) – project ID 390715994, as well as of the city of Hamburg's state-funded research project "Control of the special properties of water in nanopores" (LFF-FV68) is also acknowledged.

## Data availability

The raw data for this study were collected at the Institut Laue-Langevin (ILL) large-scale facility and the dataset will be publicly available after the embargo period (i.e., five years after conducting the experiments).[43, 44] Data derived from these findings can be obtained from the corresponding author upon request.